\documentclass[preprint,amssymb,nobibnotes,nofootinbib,aps,prc]{revtex4}
\usepackage{psfig}
\usepackage{graphicx}
\usepackage{bm}
\newcommand{\bea}{\begin{eqnarray}}
\newcommand{\eea}{\end{eqnarray}}
\newcommand{\be}{\begin{equation}}
\newcommand{\ee}{\end{equation}}
\newcommand{\bt}{\begin{tabular}}
\newcommand{\et}{\end{tabular}}
\newcommand{\Tr}{{\rm Tr}}
\newcommand{\no}{\nonumber}
\newcommand{\ovl}{\overline}

\newcommand{\Si}{ \mbox{\boldmath $\Sigma$}  }

\newcommand{\pa}{\partial}
\newcommand{\beas}{\begin{eqnarray*}}
\newcommand{\eeas}{\end{eqnarray*}}

\newcommand{\fr}{\frac}

\newcommand{\dg}{\dagger}
\newcommand{\La}{\Lambda}
\newcommand{\pam}{\partial_\mu}
\begin{document}

\title{Isospin dependent kaon and antikaon optical potentials in
dense hadronic matter}

\author{Amruta Mishra}
\email{amruta@physics.iitd.ac.in}
\affiliation{Department of Physics,Indian Institute of Technology,Delhi,
New Delhi - 110 016, India}

\author{Stefan Schramm}
\email{schramm@th.physik.uni-frankfurt.de}
\affiliation{Institut f\"ur Theoretische Physik,
     J.W. Goether Universit\"at,
        Robert Mayer Str. 8-10, D-60054 Frankfurt am Main, Germany}

\begin{abstract}
Isospin effects on the optical potentials of kaons and antikaons
in dense hadronic matter are investigated using a chiral SU(3)
model. These effects are important for asymmetric heavy ion
collision experiments. In the present work the dispersion
relations are derived for kaons and antikaons, compatible with the
low energy scattering data, within our model approach. The
relations result from the kaonic interactions with the nucleons,
vector mesons and scalar mesons in the asymmetric nuclear matter.
The isospin asymmetry effects arising from the interactions with
the vector-isovector $\rho$- meson as well as the scalar isovector
$\delta$ mesons are considered. The density dependence of the
isospin asymmetry is seen to be appreciable for the kaon and
antikaon optical potentials. This can be particularly relevant for
the future accelerator facility FAIR at GSI,
where experiments using neutron rich beams are planned to be used
in the study of compressed baryonic matter.
\end{abstract}


\maketitle

\section{Introduction}

The study of the properties of hadrons in hot and dense matter
\cite{QM02} is an important topic in the present strong
interaction physics.  This subject has direct implications for
heavy-ion collision experiments, for the study of astrophysical
compact objects (like neutron stars) as well as for the physics of
the early universe.  The in-medium properties of kaons have been
investigated particularly because of their relevance in neutron
star phenomenology as well as relativistic heavy-ion collisions.
For example, in the interior of a neutron star the attractive kaon
nucleon interaction might lead to kaon condensation as originally
suggested by Kaplan and Nelson \cite{kaplan}.  The in-medium
modification of kaon/antikaon properties can primarily be observed
experimentally in relativistic nuclear collisions. Indeed, the
experimental \cite{FOPI,Laue99,kaosnew,Sturm01,Forster02} and
theoretical studies
\cite{lix,cmko,Li2001,Cass97,brat97,CB99,laura03,Effenber00,Aichelin,Fuchs}
on $K^\pm$ production from A+A collisions at SIS energies of 1-2
A$\cdot$GeV have shown that in-medium properties of kaons have
been seen in the collective flow pattern of $K^+$ mesons as well
as in the abundance and spectra of antikaons.

The theoretical research work on the topic of medium modification
of hadron properties was initiated by Brown and Rho \cite{brown}
who suggested that the modifications of hadron masses should scale
with the scalar quark condensate $\langle q\bar{q}\rangle$ at
finite baryon density.  The first attempts to extract the
antikaon-nucleus potential from the analysis of kaonic-atom data
were in favor of very strong attractive potentials of the order of
-150 to -200 MeV at normal nuclear matter density $\rho_0$
\cite{FGB94,Gal}.  However, more recent self-consistent
calculations based on a chiral Lagrangian
\cite{Lutz98,Lutz021,Lutz02,Oset00} or coupled-channel G-matrix
theory (within meson-exchange potentials) \cite{lauran} only
predicted moderate attraction with potential depths of -50 to -80
MeV at density $\rho_0$.

The problem with the antikaon potential at finite baryon density
is that the antikaon-nucleon amplitude in the isospin channel
$I=0$ is dominated by the $\Lambda(1405)$ resonant structure,
which in free space is only 27 MeV below the ${\bar K}N$
threshold. It is presently not clear if this physical resonance is
a real excited state of a 'strange' baryon or some short
lived molecular intermediate state which can be described
in a coupled channel $T$-matrix scattering equation
using a suitable meson-baryon potential. Additionally, the
coupling between the ${\bar K}N$ and $\pi Y$ ($Y=\Lambda,\Sigma$)
channels is essential to get the proper dynamical behavior in free
space. Correspondingly, the in-medium properties of the
$\Lambda(1405)$, such as its pole position and its width, which in
turn influence strongly the antikaon-nucleus optical potential,
are very sensitive to the many-body treatment of the medium
effects. Previous works have shown that a self-consistent
treatment of the $\bar{K}$ self energy has a strong impact on the
scattering amplitudes
\cite{Lutz98,Oset00,Laura,Effenber00,Lutz02,lauran} and thus on
the in-medium properties of the antikaon.  Due to the complexity
of this many-body problem the actual kaon and antikaon self
energies (or potentials) are still a matter of debate.

The topic of isospin effects in asymmetric nuclear matter has
gained interest in the recent past \cite{asym}. The isospin
effects are important in isospin asymmetric heavy ion collision
experiments.  Within the UrQMD model the density dependence of the
symmetry potential has been studied by investigating observables
like the $\pi^-/\pi^+$ ratio, the n/p ratio \cite{li1}, the
$\Delta ^-/\Delta ^{++}$ ratio as well as the effects on the
production of $K^0$ and $K^+$ \cite {li2} and on pion flow
\cite{li3} for neutron rich heavy ion collisions. Recently, the
isospin dependence of the in-medium NN cross section \cite{li4}
has also been investigated.

In the present investigation we will use a chiral SU(3) model for
the description of hadrons in the medium \cite{paper3}.  The
nucleons -- as modified in the hot hyperonic matter -- have been
studied previously within this model \cite{kristof1}. Furthermore,
the properties of vector mesons \cite{hartree,kristof1} -- due to
their interactions  with nucleons in the medium -- have also been
examined and have been found to have appreciable modifications due
to Dirac sea polarization effects. The chiral SU(3)$_{flavor}$
model was also generalized to SU(4)$_{flavor}$ to study the mass
modification of D-mesons arising from their interactions with the
light hadrons in hot hadronic matter in \cite{dmeson}. The
energies of kaons (antikaons) at zero momentum, as modified in the
medium due to their interaction with nucleons, consistent with the
low energy KN scattering data \cite{juergen}, were also studied
within this framework \cite{kmeson1}. In the present work, we
consider the effect of isospin asymmetry on the kaon and antikaon
optical potentials in the asymmetric nuclear matter.

The outline of the paper is as follows: In section II we shall
briefly review the SU(3) model used in the present investigation.
Section III describes the medium modification of the K($\bar K$)
mesons in this effective model. In section IV, we discuss the
results obtained for the optical potentials of the kaons and
antikaons and the isospin-dependent effects on these optical
potentials in asymmetric nuclear matter. Section V summarizes the
findings of the present investigation and discusses possible
extensions of the calculations.

\section{ The hadronic chiral $SU(3) \times SU(3)$ model }
In this section the various terms of the effective hadronic Lagrangian
used
\be
{\cal L} = {\cal L}_{kin} + \sum_{ W =X,Y,V,{\cal A},u }{\cal L}_{BW}
          + {\cal L}_{VP} + {\cal L}_{vec} + {\cal L}_0 + {\cal L}_{SB}
\label{genlag}
\ee
are discussed. Eq. (\ref{genlag}) corresponds to a relativistic quantum
field theoretical model of baryons and mesons built on
a nonlinear realization of chiral symmetry and broken scale invariance
(for details see \cite{paper3,hartree,kristof1})
to describe strongly interacting nuclear matter.
The model was used successfully to describe nuclear matter, finite nuclei,
hypernuclei and neutron stars.
The Lagrangian contains the baryon octet, the spin-0 and spin-1 meson
multiplets as the elementary degrees of freedom. In Eq. (\ref{genlag}),
$ {\cal L}_{kin} $ is the kinetic energy term, $  {\cal L}_{BW}  $
contains the baryon-meson interactions in which the baryon-spin-0 meson
interaction terms generate the baryon masses. $ {\cal L}_{VP} $
describes the interactions of vector mesons with the pseudoscalar
mesons (and with photons).  $ {\cal L}_{vec} $ describes the dynamical
mass generation of the vector mesons via couplings to the scalar
mesons and contains additionally quartic self-interactions of the
vector fields.  ${\cal L}_0 $ contains the meson-meson interaction terms
inducing the spontaneous breaking of chiral symmetry as well as
a scale invariance breaking logarithmic potential. $ {\cal L}_{SB} $
describes the explicit chiral symmetry breaking.

The kinetic energy terms are given as
\bea
\label{kinetic}
{\cal L}_{kin} &=& i\Tr \overline{B} \gamma_{\mu} D^{\mu}B
                + \frac{1}{2} \Tr D_{\mu} X D^{\mu} X
+  \Tr (u_{\mu} X u^{\mu}X +X u_{\mu} u^{\mu} X)
                + \frac{1}{2}\Tr D_{\mu} Y D^{\mu} Y \nonumber \\
               &+&\frac {1}{2} D_{\mu} \chi D^{\mu} \chi
                - \frac{ 1 }{ 4 } \Tr
\left(\tilde V_{ \mu \nu } \tilde V^{\mu \nu }  \right)
- \frac{ 1 }{ 4 } \Tr \left(F_{ \mu \nu } F^{\mu \nu }  \right)
- \frac{ 1 }{ 4 } \Tr \left( {\cal A}_{ \mu \nu } {\cal A}^{\mu \nu }
 \right)\, .
\eea
In (\ref{kinetic}) $B$ is the baryon octet, $X$ the scalar meson
multiplet, $Y$ the pseudoscalar chiral singlet, $\tilde{V}^\mu$ (${\cal
A}^\mu$) the renormalised vector (axial vector) meson multiplet with
the field strength tensor
$\tilde{V}_{\mu\nu}=\pa_\mu\tilde{V}_\nu-\pa_\nu\tilde{V}_\mu$ $({\cal
A}_{\mu\nu}= \pa_\mu{\cal A}_\nu-\pa_\nu{\cal A}_\mu $), $F_{\mu\nu}$
is the field strength tensor of the photon and $\chi$
is the scalar, iso-scalar dilaton (glueball) -field.
In the above, $u_\mu= -\fr{i}{2}[u^\dg\pam u - u\pam u^\dg]$,
where $u=\exp\Bigg[\fr{i}{\sigma_0}\pi^a\lambda^a\gamma_5\Bigg]$
is the unitary transformation operator, and the covariant derivative
reads $ D_\mu = \pam\, + [\Gamma_\mu,\,\,]$, with
$\Gamma_\mu=-\fr{i}{2}[u^\dg\pam u + u\pam u^\dg]$.

The baryon -meson interaction for a general meson field $W$ has
the form \be {\cal L}_{BW} = -\sqrt{2}g_8^W
\left(\alpha_W[\ovl{B}{\cal O}BW]_F+ (1-\alpha_W) [\ovl{B} {\cal
O}B W]_D \right) - g_1^W \frac{1}{\sqrt{3}} \Tr(\ovl{B}{\cal O}
B)\Tr W  \, , \ee with $[\ovl{B}{\cal O}BW]_F:=\Tr(\ovl{B}{\cal
O}WB-\ovl{B}{\cal O}BW)$ and $[\ovl{B}{\cal O}BW]_D:=
\Tr(\ovl{B}{\cal O}WB+\ovl{B}{\cal O}BW) - \frac{2}{3}\Tr
(\ovl{B}{\cal O} B) \Tr W$. The different terms to be considered
are those for the interaction of baryons  with scalar mesons
($W=X, {\cal O}=1$), with vector mesons  ($W=\tilde V_{\mu}, {\cal
O}=\gamma_{\mu}$ for the vector and $W=\tilde V_{\mu \nu}, {\cal
O}=\sigma^{\mu \nu}$ for the tensor interaction), with axial
vector mesons ($W={\cal A}_\mu, {\cal O}=\gamma_\mu \gamma_5$) and
with pseudoscalar mesons ($W=u_{\mu},{\cal
O}=\gamma_{\mu}\gamma_5$), respectively. For the current
investigation the following interactions are relevant:
Baryon-scalar meson interactions generate the baryon masses
through coupling of the baryons to the non-strange $ \sigma (\sim
\langle\bar{u}u + \bar{d}d\rangle) $ and the strange $
\zeta(\sim\langle\bar{s}s\rangle) $ scalar quark condensate. The
parameters $ g_1^S$, $g_8^S $ and $\alpha_S$ are adjusted to fix
the baryon masses to their experimentally measured vacuum values.
It should be emphasized that the nucleon mass also depends on the
{\em strange condensate} $ \zeta $. For the special case of ideal
mixing ($\alpha_S=1$ and $g_1^S=\sqrt 6 g_8^S$) the nucleon mass
depends only on the non--strange quark condensate. In the present
investigation, the general case will be used, which takes into account
the baryon coupling terms to both scalar fields ($\sigma$ and
$\zeta$).

In analogy to the baryon-scalar meson coupling two
independent baryon-vector meson interaction terms exist corresponding to
the $F$-type (antisymmetric) and $D$-type (symmetric) couplings.
Here we will use the antisymmetric coupling because, following the
universality principle  \cite{saku69} and the vector meson
dominance model, one can conclude that the symmetric coupling
should be small. We realize it by setting $\alpha_V=1$ for all
fits. Additionally we decouple the strange vector field $
\phi_\mu\sim\bar{s} \gamma_\mu s $ from the nucleon by setting $
g_1^V=\sqrt{6}g_8^V $. The remaining baryon-vector meson
interaction reads \be {\cal
L}_{BV}=-\sqrt{2}g_8^V\Big\{[\bar{B}\gamma_\mu
BV^\mu]_F+\Tr\big(\bar{B}\gamma_\mu B\big) \Tr V^\mu\Big\}\, . \ee

The Lagrangian describing the interaction for the scalar mesons, $X$,
and pseudoscalar singlet, $Y$, is given as \cite{paper3}
\bea
\label{cpot}
{\cal L}_0 &= &  -\frac{ 1 }{ 2 } k_0 \chi^2 I_2
     + k_1 (I_2)^2 + k_2 I_4 +2 k_3 \chi I_3,
\eea with $I_2= \Tr (X+iY)^2$, $I_3=\det (X+iY)$ and $I_4 = \Tr
(X+iY)^4$. In the above, $\chi$ is the scalar color singlet
glueball field. It is introduced in order to mimic the QCD trace
anomaly, i.e. the non-vanishing energy-momentum tensor
$\theta_\mu^\mu = (\beta_{QCD}/2g)\langle
G^a_{\mu\nu}G^{a,\mu\nu}\rangle$, where $G^a_{\mu\nu}$ is the
gluon field tensor. A scale breaking potential is introduced: \be
\label{lscale}
  {\cal L}_{\mathrm{scalebreak}}=- \frac{1}{4}\chi^4 \ln
   \frac{ \chi^4 }{ \chi_0^4}
 +\frac{\delta}{3}\chi^4 \ln \frac{I_3}{\det \langle X \rangle_0}
\ee
which allows for the identification of the $\chi$ field width the gluon
condensate $\theta_\mu^\mu=(1-\delta)\chi^4$.
Finally the term
${\cal L}_{\chi} = - k_4 \chi^4 $
generates a phenomenologically consistent finite vacuum expectation
value. The variation of
$\chi$ in the medium is rather small \cite{paper3}.
Hence we shall use the frozen glueball approximation i.e. set
$\chi$ to its vacuum value, $\chi_0$.

The Lagrangian for the vector meson interaction is written as
\bea
{\cal L}_{vec} &=&
    \fr{m_V^2}{2}\fr{\chi^2}{\chi_0^2}\Tr\big(\tilde{V}_\mu\tilde{V}^\mu\big)
+   \fr{\mu}{4}\Tr\big(\tilde{V}_{\mu\nu}\tilde{V}^{\mu\nu}X^2\big) 
+ \fr{\lambda_V}{12}\Big(\Tr\big(\tilde{V}^{\mu\nu}\big)\Big)^2 +
    2(\tilde{g}_4)^4\Tr\big(\tilde{V}_\mu\tilde{V}^\mu\big)^2  \, .
\eea
The vector meson fields, $\tilde{V}_\mu$ are related to the
renormalized fields by
$V_\mu = Z_V^{1/2}\tilde{V}_\mu$, with $V = \omega, \rho, \phi \, $.
The masses of $\omega,\rho$ and $\phi$ are fitted from $m_V, \mu$ and
$\lambda_V$.

The explicit symmetry breaking term is given as \cite{paper3}
\be
 {\cal L}_{SB}=\Tr A_p\left(u(X+iY)u+u^\dagger(X-iY)u^\dagger\right)
\label{esb-gl}
\ee
with $A_p=1/\sqrt{2}{\mathrm{diag}}(m_{\pi}^2 f_{\pi},m_\pi^2 f_\pi, 2 m_K^2 f_K
-m_{\pi}^2 f_\pi)$ and $m_{\pi}=139$ MeV, $m_K=498$ MeV. This
choice for $A_p$, together with the constraints
$\sigma_0=-f_\pi$, $\zeta_0=-\frac {1}{\sqrt 2} (2 f_K -f_\pi)$
for the VEV of the scalar condensates assure that
the PCAC-relations of the pion and kaon are fulfilled.
With $f_{\pi} = 93.3$~MeV and $f_K = 122$~MeV we obtain $|\sigma_0| =
93.3$~MeV and $|\zeta_0 |= 106.56$~MeV.

We proceed to study the hadronic properties in the chiral SU(3) model.
The Lagrangian density in the mean field approximation is given as
\begin{eqnarray}
{\cal L}_{BX}+{\cal L}_{BV} &=& -\sum_i\overline{\psi_{i}}\, [g_{i
\omega}\gamma_0 \omega + g_{i\phi}\gamma_0 \phi
+m_i^{\ast} ]\,\psi_{i} \\
{\cal L}_{vec} &=& \frac{1}{2}m_{\omega}^{2}\frac{\chi^2}{\chi_0^2}\omega^
2+g_4^4 \omega^4 +
\frac{1}{2}m_{\phi}^{2}\frac{\chi^2}{\chi_0^2}\phi^2+g_4^4
\left(\fr{Z_\phi}{Z_\omega}\right)^2\phi^4\\
{\cal V}_0 &=& \frac{ 1 }{ 2 } k_0 \chi^2 (\sigma^2+\zeta^2)
- k_1 (\sigma^2+\zeta^2)^2
     - k_2 ( \frac{ \sigma^4}{ 2 } + \zeta^4)
     - k_3 \chi \sigma^2 \zeta \nonumber \\
&+& k_4 \chi^4 + \frac{1}{4}\chi^4 \ln \frac{ \chi^4 }{ \chi_0^4}
 -\frac{\delta}{3} \chi^4 \ln \frac{\sigma^2\zeta}{\sigma_0^2 \zeta_0} \\
{\cal V}_{SB} &=& \left(\frac{\chi}{\chi_0}\right)^{2}\left[m_{\pi}^2 f_{\pi}
\sigma
+ (\sqrt{2}m_K^2 f_K - \frac{ 1 }{ \sqrt{2} } m_{\pi}^2 f_{\pi})\zeta
\right],
\end{eqnarray}
where $m_i^* = -g_{\sigma i}{\sigma}-g_{\zeta i}{\zeta} $ is the
effective mass of the baryon of type i ($i = N, \Si, \La, \Xi $).
In the above, $g_4=\sqrt {Z_\omega} \tilde g_4$ is the
renormalized coupling for $\omega$-field. The thermodynamical
potential of the grand canonical ensemble $\Omega$ per unit volume
$V$ at given chemical potential $\mu$ and temperature $T$ can be
written as \bea \frac{\Omega}{V} &=& -{\cal L}_{vec} - {\cal L}_0
- {\cal L}_{SB} - {\cal V}_{vac} + \sum_i\frac{\gamma_i }{(2
\pi)^3} \int d^3k\, E^{\ast}_i(k)\Big(f_i(k)+\bar{f}_i(k)
\Big)\nonumber \\
&&- \sum_i\frac{\gamma_i }{(2 \pi)^3}\,\mu^{\ast}_i
\int d^3k\,\Big(f_i(k)-\bar{f}_i(k)\Big).
\label{OmegaV}
\eea
Here the the potential at $\rho=0$ has been subtracted
in order to get a vanishing vacuum energy. In (\ref{OmegaV}) $\gamma_i$
are the spin-isospin degeneracy factors.  The $f_i$ and $\bar{f}_i$ are
thermal distribution functions for the baryon of species $i$, given in
terms of the effective single particle energy, $E^\ast_i$, and chemical
potential, $\mu^\ast_i$, as
\bea
f_i(k) &=& \fr{1}{{\rm e}^{\beta (E^{\ast}_i(k)-\mu^{\ast}_i)}+1}\quad ,\quad
\bar{f}_i(k)=\fr{1}{{\rm e}^{\beta (E^{\ast}_i(k)+\mu^{\ast}_i)}+1}, \no
\eea
with $E^{\ast}_i(k) = \sqrt{k_i^2+{m^\ast_i}^2}$ and $ \mu^{\ast}_i
                = \mu_i-g_{i\omega}\omega$.
The mesonic field equations are determined by minimizing the
thermodynamical potential \cite{hartree,kristof1}.
They depend on
the scalar and vector densities
for the baryons at finite temperature
\bea
\rho^s_i = \gamma_i
\int \frac{d^3 k}{(2 \pi)^3} \,\frac{m_i^{\ast}}{E^{\ast}_i}\,
\left(f_i(k) + \bar{f}_i(k)\right) \, ; \;\;
\rho_i = \gamma_i \int \frac{d^3 k}{(2 \pi)^3}\,\left(f_i(k) -
\bar{f}_i(k)\right) \,.
\label{dens}
\eea
The energy density and the pressure are given as,
$\epsilon = \Omega/V+\mu_i\rho_i $+TS and $ p = -\Omega/V $.

\begin{figure}
\phantom{a}\hspace*{-2cm}
\psfig{file=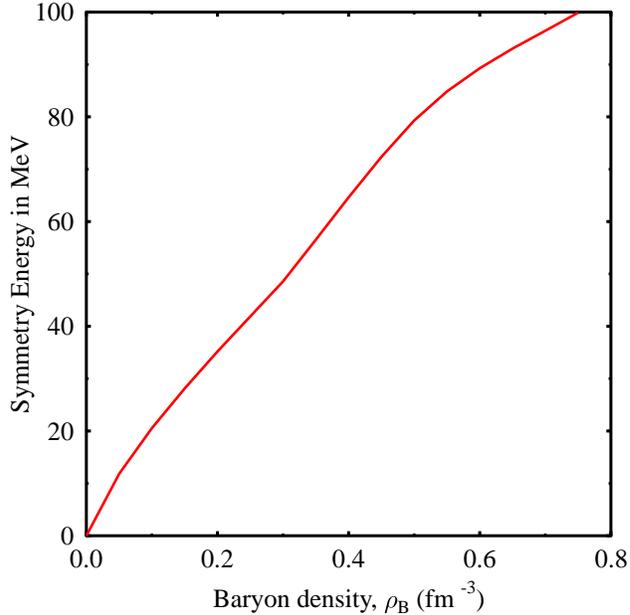,width=11cm}
\caption{
The symmetry energy in MeV plotted as a function of the baryon
density, $\rho_B$ (in fm $^{-3}$).
}
\label{asymdens}
\end{figure}

\section{Kaon (antikaon) interactions in the chiral SU(3) model}
\label{kmeson}

In this section, we derive the disperson relations
for the $K (\bar K)$ and calculate their optical potentials
in the asymmetric nuclear matter. The medium modified
energies of the kaons and antikaons arise from their interactions
with the nucleons, vector mesons and scalar mesons
within the chiral SU(3) model.

In this model
the interactions with the scalar fields (non-strange, $\sigma$ and strange,
$\zeta$), scalar--isovector field $\delta$ as well as a vectorial
interaction and the vector meson ($\omega$ and $\rho$) - exchange
terms modify the energies for $K(\bar K)$ mesons in the medium.
In the following, we shall derive the dispersion relations
for the kaons and antikaons, including the effects from isospin
asymmetry originating from both the vector-isovector $\rho$-field
as well as the scalar-isovector $\delta$ field.

\begin{figure}
\phantom{a}\hspace*{-2cm}
\psfig{file=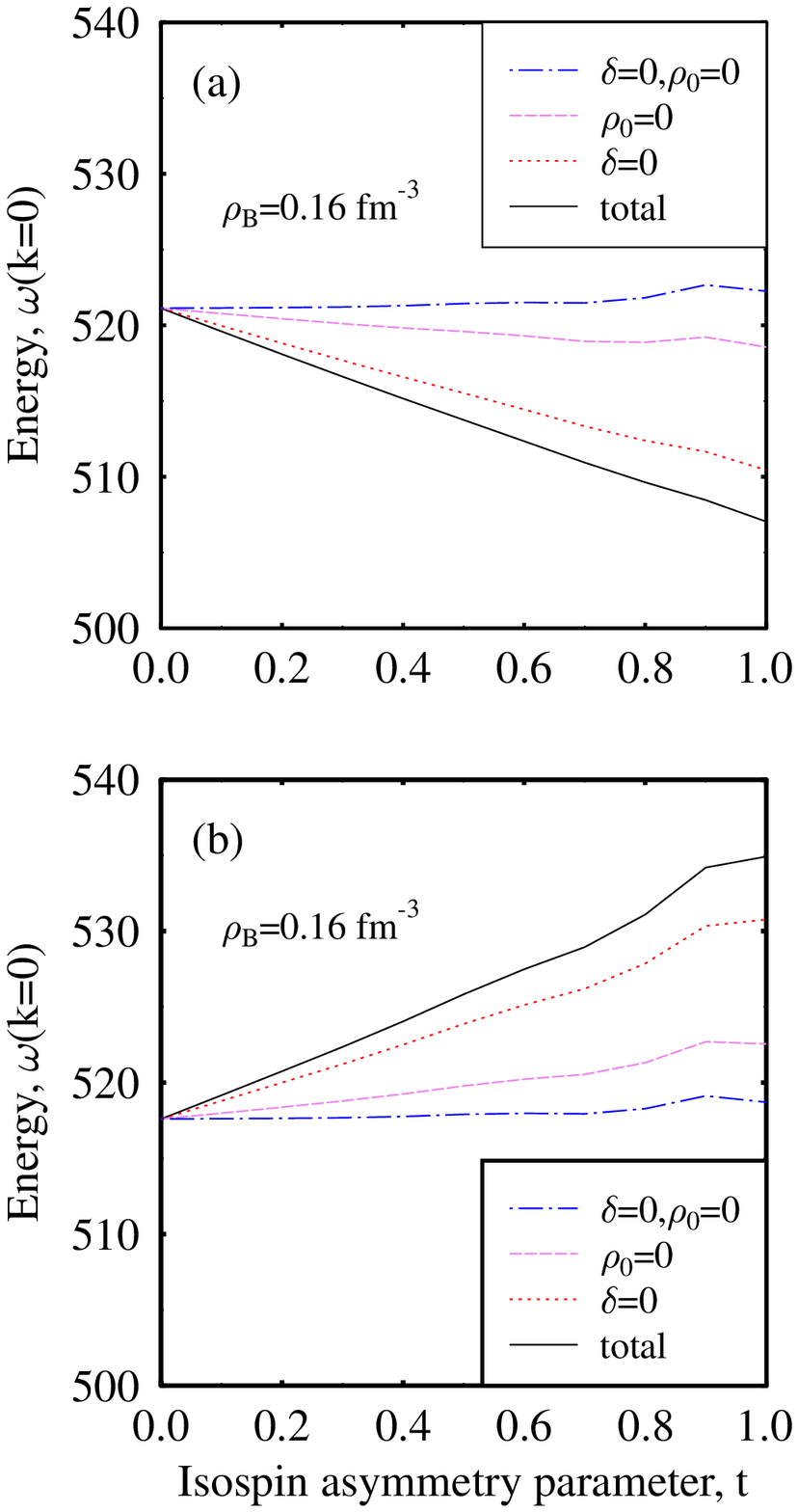,width=16cm}
\caption{
The energies of the kaons, $K^+$ and $K^0$, at zero momentum
and for $\rho_B$=0.16 fm$^{-3}$,
are plotted as functions of the isospin asymmetry parameter, t
in (a) and (b).
The medium modifications to the energies are also shown for the
situations when either the isospin asymmetric contribution from
the $\rho$-meson or $\delta$ meson, or, both,
are not taken into account. The solid line shows the total contribution.
}
\label{mk16t}
\end{figure}

\begin{figure}
\phantom{a}\hspace*{-2cm}
\psfig{file=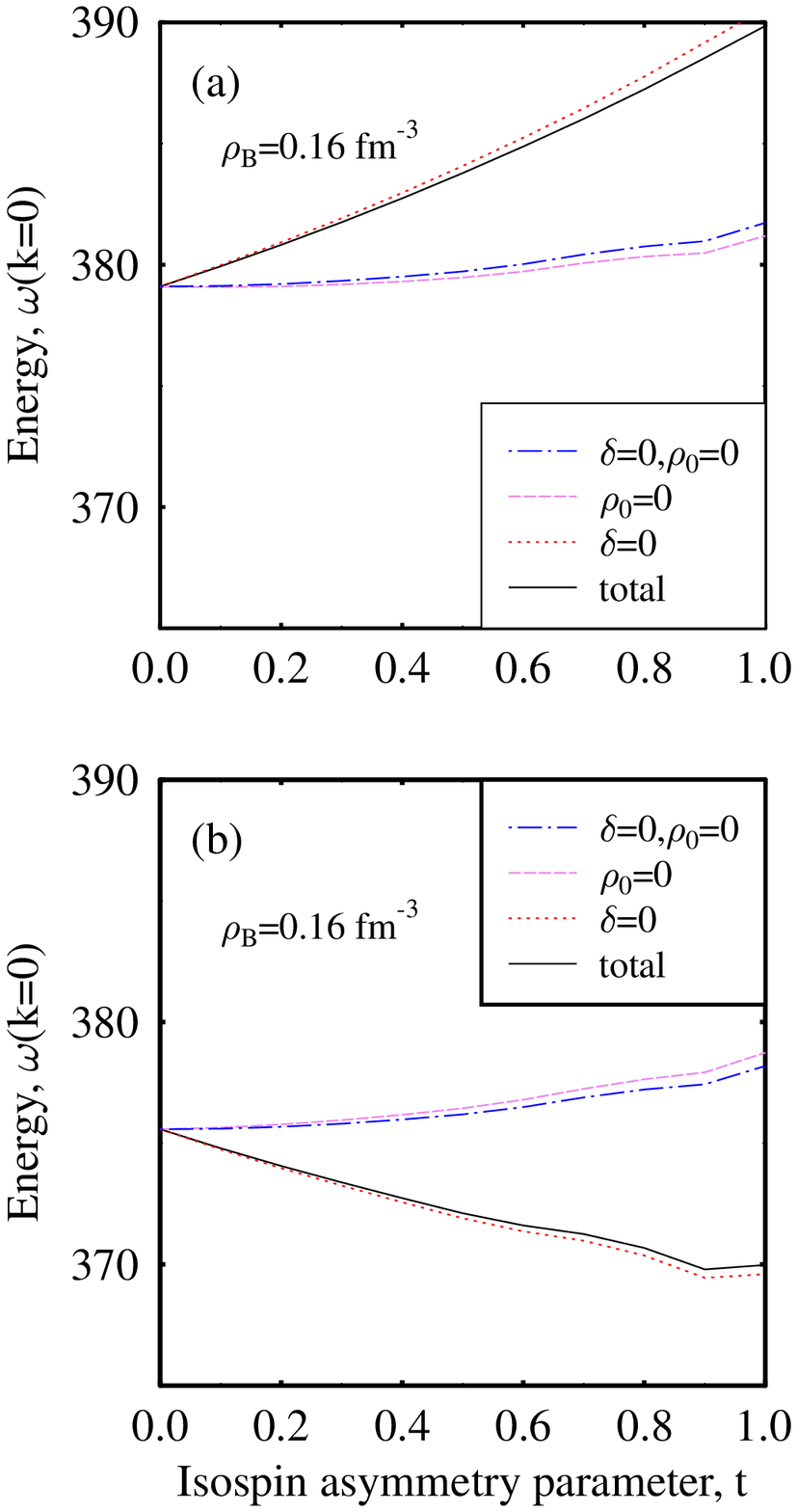,width=16cm}
\caption{
The energies of the antikaons, $K^-$ and ${\bar {K^0}}$, at zero momentum
and for $\rho_B$=0.16 fm$^{-3}$ are plotted as functions of the
the isospin asymmetry parameter, t in (a) and (b).
The medium modifications to the energies are also shown for the
situations when the isospin asymmetric contribution from
the $\rho$-meson or $\delta$ meson, or, both,
are not taken into account. The solid line shows the total contribution.
The symmetry energy in MeV plotted as a function of the baryon
density, $\rho_B$ (in fm $^{-3}$).
}
\label{mkbmft}
\end{figure}

The scalar meson multiplet has the expectation value
$\langle X \rangle
= \rm {diag}( (\sigma+\delta)/\sqrt 2, (\sigma-\delta)/\sqrt 2 , \zeta ) $,
with $\sigma$ ans $\zeta$ corresponding to the non-strange and strange
scalar condensates, and $\delta$ is the third isospin component of the
scalar-isovector field, $\vec \delta$.
The pseudoscalar meson field $P$ can be written as,
\begin{equation}
P = \left(
\begin{array}{ccc} \pi^0/\sqrt 2 & \pi^+ & \frac{2 K^+}{1+w} \\
\pi^- & -\pi^0/\sqrt 2 & \frac {2 K^0}{1+w} \\
\frac {2 K^-}{1+w} & \frac {2 \bar {K^0}}{1+w} & 0
\\ \end{array}\right),
\end{equation}
where $w=\sqrt 2 \zeta/\sigma$ and we have written down
the terms that are relevant for the present investigation.
From PCAC one gets the decay constants for the pseudoscalar mesons
as $f_\pi=-\sigma$ and $f_K=-(\sigma +\sqrt 2 \zeta )/2$.
The vector meson interaction with
the pseudoscalar mesons, which modifies the energies of the K
($\bar {rm K}$ mesons,
is given as \cite {kmeson}
\begin{equation}
{\cal L} _ {VP}= -\frac{m_V^2}{2g_V} {\rm {Tr}} (\Gamma_\mu V^\mu) +
{\rm  h.c.}
\label{lvp}
\end{equation}
The vector meson multiplet is given as
$V = {\rm  diag}\big ((\omega +\rho_0)/\sqrt 2,\;
(\omega -\rho_0)/\sqrt 2,\; \phi\big )$. The non-diagonal
components in the multiplet, which are not relevant in the present
investigation, are omitted. With the interaction (\ref{lvp}),
the coupling of the $K$-meson to the $\omega$-meson is related to  the
pion-rho coupling as
$g_{\omega K}/g_{\rho \pi \pi}=f_\pi^2 /(2f_K^2)$.

The scalar meson exchange interaction term
is determined from the explicit symmetry breaking term
by equation (\ref {esb-gl}), where $A_p =1/\sqrt 2$
diag ($m_\pi^2 f_\pi$, $m_\pi^2 f_\pi$, 2 $m_K^2 f_K -m_\pi^2 f_\pi$).

The interaction Lagrangian modifying the energies of the $K(\bar K)$-mesons
can be written as
\begin{eqnarray}
\cal L _{KN} & = & -\frac {i}{8 f_K^2} \Big [ 3 ( \bar N \gamma^\mu N)
(\bar K (\partial_\mu K) - (\partial_\mu {\bar K})  K)
+(\bar N \gamma^\mu \tau^a N)
(\bar K \tau^a (\partial_\mu  K) - (\partial_\mu {\bar K})\tau ^a  K)\Big ]
\nonumber \\
 &+ & \frac{m_K^2}{2f_K} \Big [ (\sigma +\sqrt 2 \zeta)(\bar  K K)
+\delta ^a (\bar  K \tau^a K)\Big ] \nonumber \\
 &- & i g_{\omega K}\Big [ ({\bar  K} (\partial_\mu K)
- (\partial_\mu {\bar K}) K) \omega ^ \mu
 +({\bar  K} \tau^a (\partial_\mu  K)
- (\partial_\mu {\bar K})\tau^a K) {\rho ^ \mu}^a\Big ]
 \nonumber \\
& - & \frac {1}{f_K}\Big [ (\sigma +\sqrt 2 \zeta)
(\partial _\mu {\bar  K})(\partial ^\mu K)
+(\partial _\mu  {\bar  K})\tau^a(\partial ^\mu K)\delta ^a \Big ]
\nonumber \\
&+ &\frac {d_1}{2 f_K^2}(\bar N N)
(\partial _\mu {\bar K})(\partial ^\mu K).
\label{lagd}
\end{eqnarray}

\begin{figure}
\phantom{a}\hspace*{-2cm}
\psfig{file=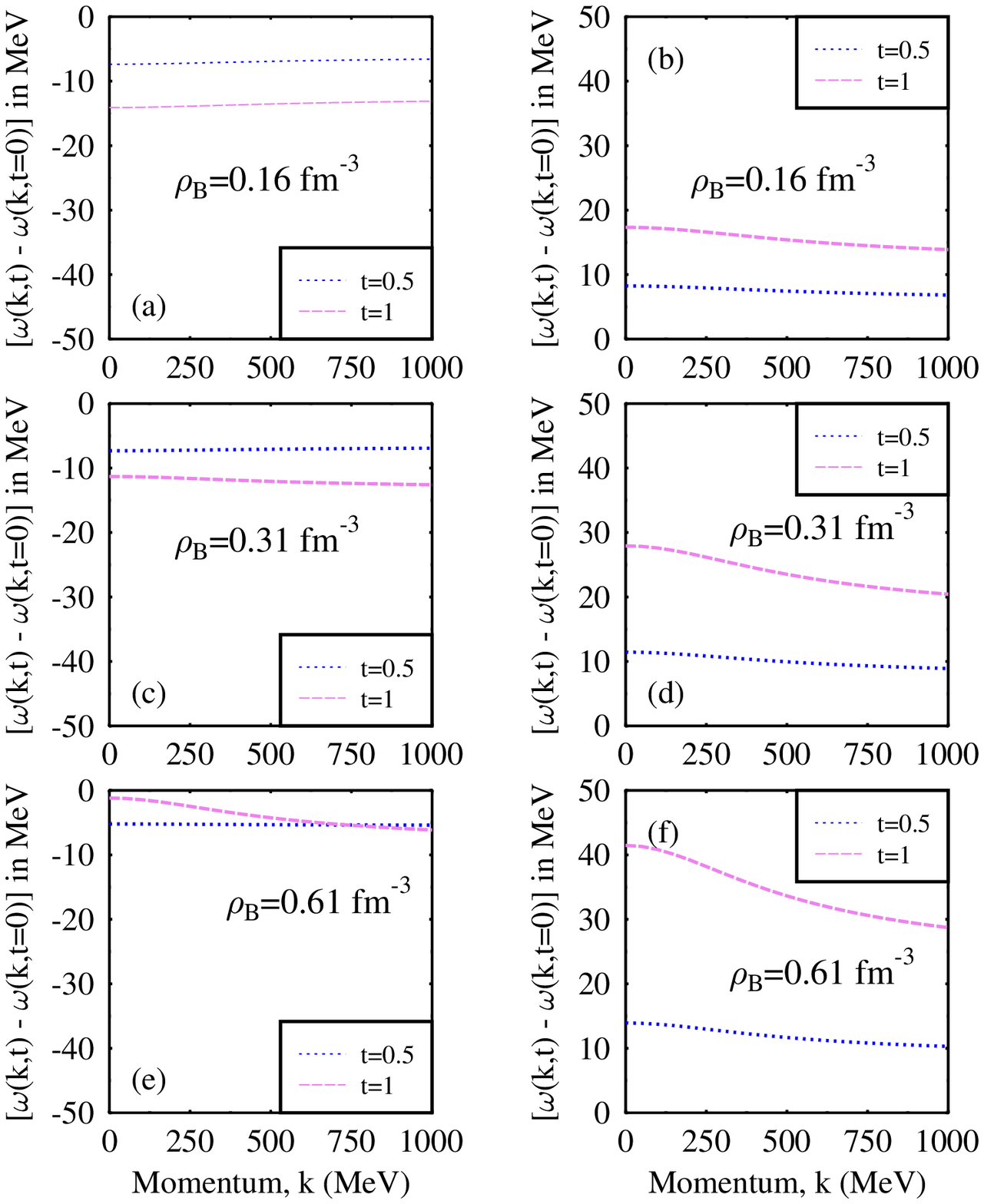,width=18cm}
\caption{
The energies of the kaons, relative to the t=0 values,
are plotted as functions of the momentum for t=0.5 and 1.
The subplots (a),(c) and (e) refer to $K^+$ and,
(b), (d) and (f) refer to $K^0$ at different densities.
}
\label{omgkp}
\end{figure}

\begin{figure}
\phantom{a}\hspace*{-2cm}
\psfig{file=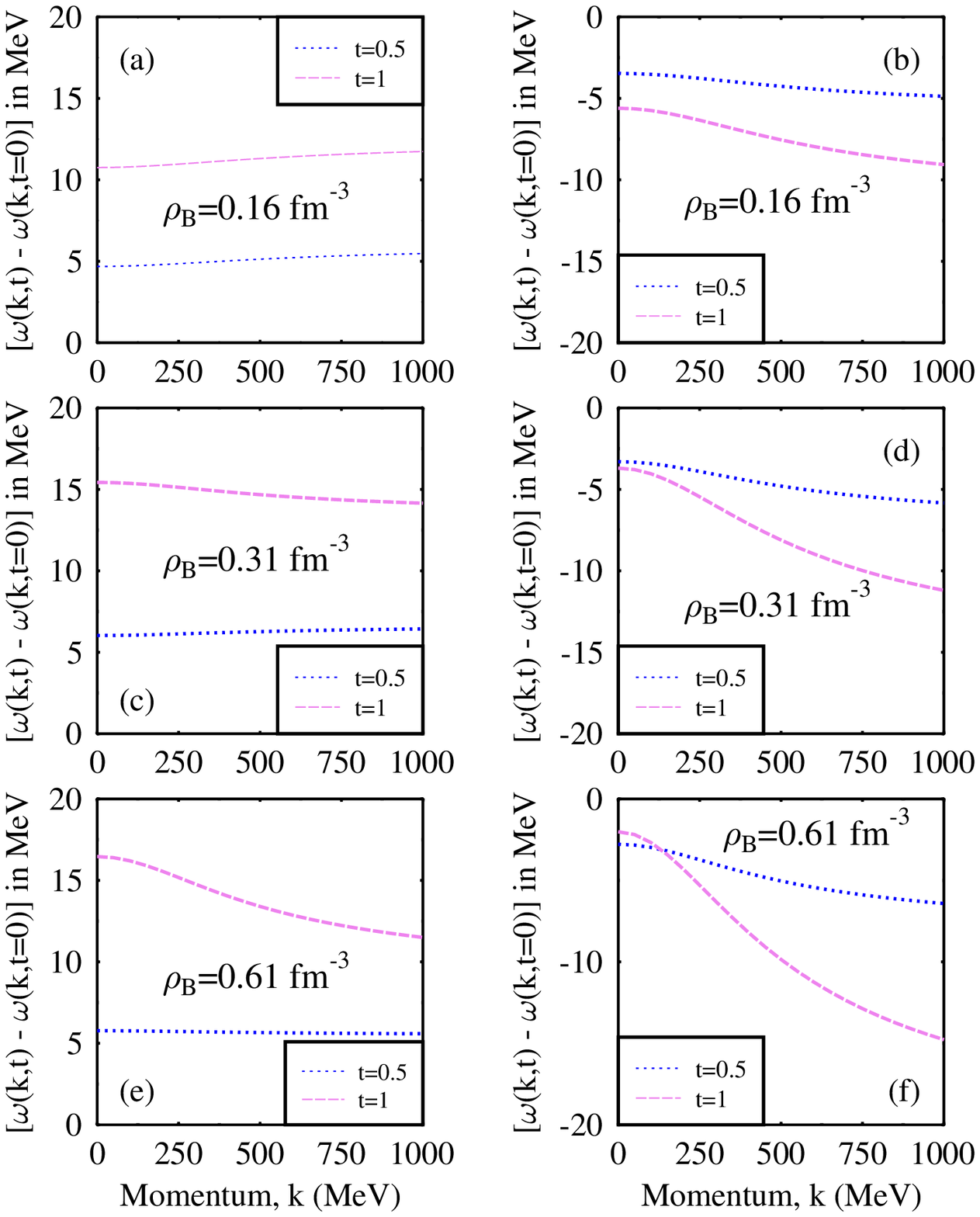,width=18cm}
\caption{
The energies of the antikaons, relative to the t=0 values,
are plotted as functions of the momentum for t=0.5 and 1.
The subplots (a),(c) and (e) refer to $K^-$ and,
(b), (d) and (f) refer to ${\bar K}^0$ at different densities.
}
\label{omgkbp}
\end{figure}

In the above, $K$ and $\bar K$ are the kaon and antikaon doublets.
In (\ref{lagd}) the first line is the vectorial interaction term
obtained from the first term in (\ref{kinetic}) (Weinberg-Tomozawa
term).  The second term, which gives an attractive interaction for the
$K$-mesons, is obtained from the explicit symmetry breaking term
(\ref{esb-gl}). The third term, arising from equation (\ref{lvp}),
 refers to the interaction in terms
of the $\omega$-meson and $\rho$-meson exchanges.
The fourth term arises within the present chiral model from
the kinetic term of the pseudoscalar mesons given by the third term in
equation (\ref{kinetic}), when the scalar fields in one of the meson
multiplets, $X$, are replaced by their vacuum expectation values.  The
fifth term in (\ref{lagd}) for the KN interactions arises from the term
\begin{equation}
{\cal L }^{BM} =d_1 Tr (u_\mu u ^\mu \bar B B),
\label{dtld}
\end{equation}
in the SU(3) chiral model \cite{kmeson1}. The last two terms in (\ref{lagd})
represent the range term in the chiral model.
The Fourier transformation of the equation of motion for kaons
(antikaons) leads to the dispersion relations,
$$-\omega^2+ {\vec k}^2 + m_K^2 -\Pi_K(\omega, |\vec k|,\rho)=0,$$
where $\Pi_K$ denotes the kaon (antikaon) self energy in the medium.

Explicitly, the self energy $\Pi (\omega,|\vec k|)$ for the kaon doublet
arising from the interaction (\ref{lagd}) is given as

\begin{eqnarray}
\Pi (\omega, |\vec k|) &= & -\frac {3}{4 f_K^2} (\rho_p +\rho_n) \omega
+\frac {m_K^2}{2 f_K} (\sigma ' +\sqrt 2 \zeta ' \pm \delta ')
-  2 g_{\omega K} \omega (\omega_0 \pm \rho_0)
\nonumber \\ & +& \Big [- \frac {1}{f_K}
(\sigma ' +\sqrt 2 \zeta ' \pm \delta ')
+\frac {d_1}{2 f_K ^2} (\rho_s ^p +\rho_s ^n) \Big ]
(\omega ^2 - {\vec k}^2),
\label{selfk}
\end{eqnarray}
where the $\pm$ signs refer to the $K^+$ and $K^0$ respectively.
In the above, $\sigma ' (=\sigma -\sigma _0)$,
$\zeta ' (=\zeta -\zeta _0)$ and  $\delta ' (=\delta -\delta _0)$,
are the fluctuations of the scalar-isoscalar fields $\sigma$ and $\zeta$,
and the third component of the scalar-isovector field, $\delta$,
from their vacuum expectation values.
The vacuum expectation value of $\delta$ is zero ($\delta_0$=0), since
a nonzero value for it will break the isospin symmetry of the vacuum
(the small isospin breaking effect arising from the mass and charge difference of the up and
down quarks, respectively,
has been neglected here).
$\rho_p$ and $\rho_n$ are the number densities
for the proton and the neutron, $\rho_s^p$
and $\rho_s^n$ are their corresponding scalar densities.

\begin{figure}
\phantom{a}\hspace*{-2cm}
\psfig{file=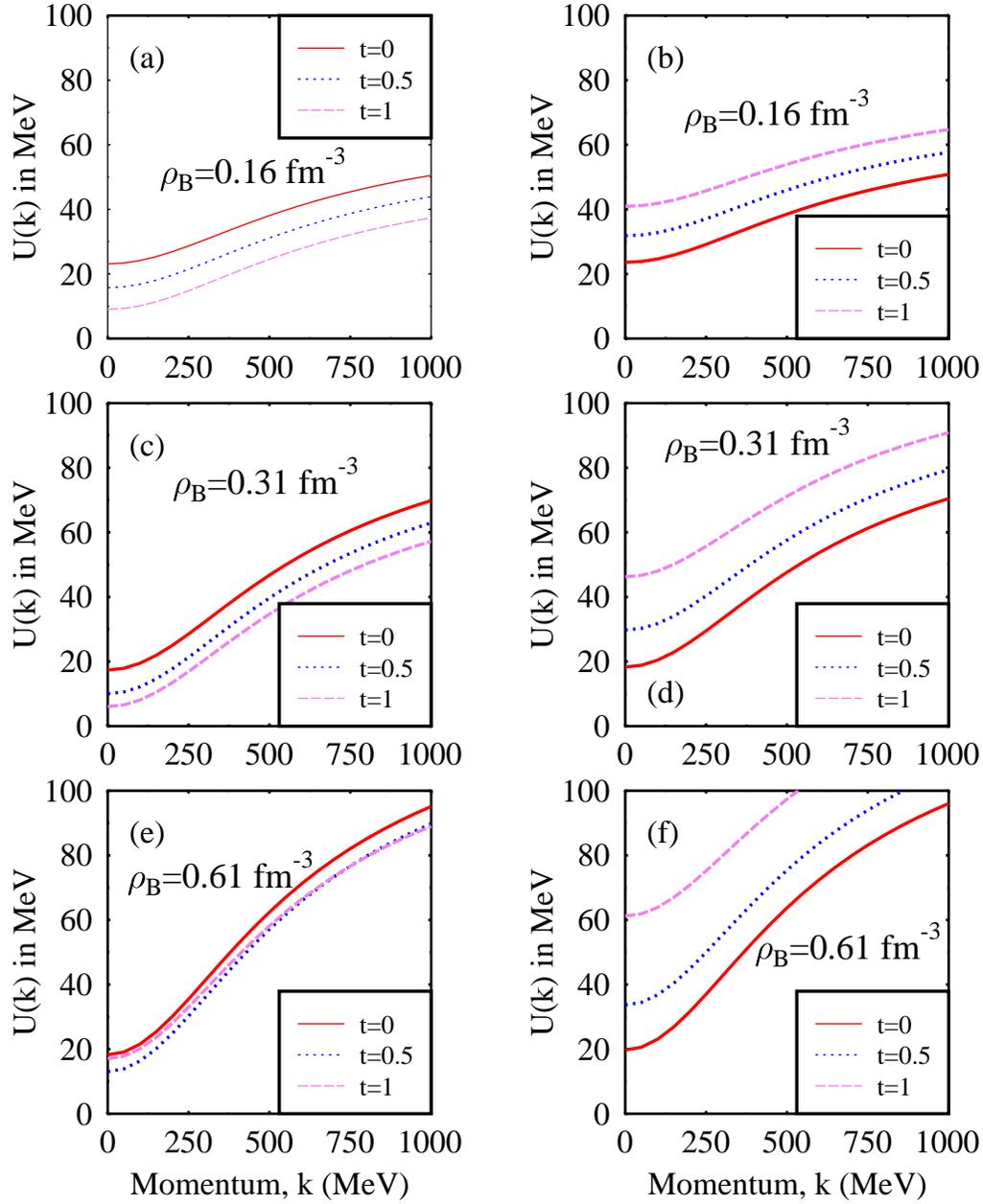,width=18cm}
\caption{
The optical potentials of the kaons
are plotted as functions of the momentum for
various values of the isospin asymmetry parameter t.
The subplots (a),(c) and (e) refer to $K^+$ and,
(b), (d) and (f) refer to $K^0$ at different densities.
}
\label{optkp}
\end{figure}

\begin{figure}
\phantom{a}\hspace*{-2cm}
\psfig{file=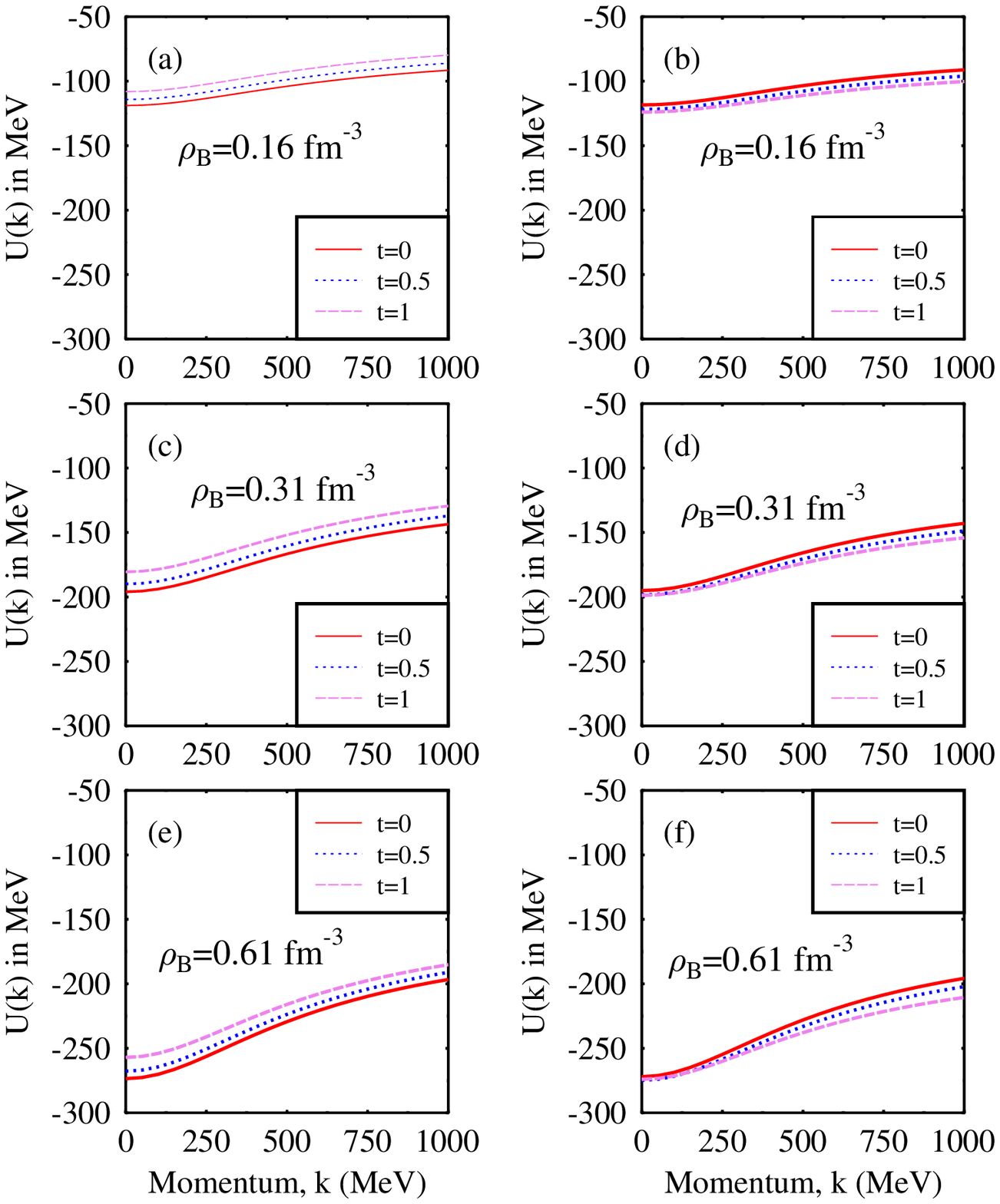,width=18cm}
\caption{
The optical potentials of the antikaons
are plotted as functions of the momentum for
various values of the isospin asymmetry parameter, t.
The subplots (a),(c) and (e) refer to $K^-$ and,
(b), (d) and (f) refer to ${\bar K}^0$ at different densities.
}
\label{optkbp}
\end{figure}

Similarly, for the antikaon doublet, the self-energy is calculated as
\begin{eqnarray}
\Pi (\omega, |\vec k|) &= & \frac {3}{4 f_K^2} (\rho_p +\rho_n) \omega
+\frac {m_K^2}{2 f_K} (\sigma ' +\sqrt 2 \zeta ' \pm \delta ')
+  2 g_{\omega K} \omega (\omega_0 \pm \rho_0)
\nonumber \\ & +& \Big [- \frac {1}{f_K}
(\sigma ' +\sqrt 2 \zeta ' \pm \delta ')
+\frac {d_1}{2 f_K ^2} (\rho_s ^p +\rho_s ^n) \Big ]
(\omega ^2 - {\vec k}^2),
\label{selfkb}
\end{eqnarray}
where the $\pm$ signs refer to the $K^-$ and $\bar {K^0}$ respectively.

After solving the above dispersion relations
for the kaons and antikaons, their optical potentials
can be calculated from

\be
U(\omega, k) = \omega (k) -\sqrt {k^2 + m_K ^2},
\ee
where $m_K$ is the vacuum mass for the kaon (antikaon).

The parameter $d_1$ is calculated from the
the empirical value of the isospin averaged KN scattering length
\cite{thorsson,juergen,barnes} taken to be
\be
\bar a _{KN} \approx -0.255 ~ \rm {fm}.
\label{aknemp}
\ee

\section{Results and Discussions}
\label{kmass}

The present calculation uses the model parameters from \cite{paper3}. The values,
$g_{\sigma N}=10.623,\;\; {\rm and}\;\; g_{\zeta N}=-0.4894$ are
determined by fitting vacuum baryon masses.
The other parameters as fitted to the nuclear matter
saturation properties in the mean field approximation are:
$g_{\omega N}$=13.606, $g_4$=61.466,
$m_\zeta$ =1038.5 MeV, $m_\sigma$= 474.3 MeV.
The coefficient $d_1$, calculated from the empirical
value of the isospin averaged scattering length (\ref{aknemp}),
is $5.196/{m_K}$. Using these parameters, the symmetry energy defined as
\be
a_4 = \frac{1}{2} \frac{d^2 E}{dt ^2}|_{t=0}
\ee
with the asymmetry parameter $t = (\rho_n-\rho_p)/\rho_B$,
has a value of $a_4 =$ 28.4 MeV at saturation nuclear matter
density of $\rho_0$=0.15 fm $^{-3}$. Figure 1 shows the density
dependence of the symmetry energy, which increases with density
similar to previous calculations \cite{asymdens}.

The kaon and antikaon properties were studied
in the isospin symmetric hadronic matter
within the chiral SU(3) model in ref. \cite{kmeson1}.
The contributions from the vector interaction as well as
the vector meson $\omega$- exchange
terms lead to a drop for the antikaons
energy, whereas they are repulsive for the kaons. The scalar meson
exchange term arising from the scalar-isoscalar fields ($\sigma$
and $\zeta$) is attractive for both $K$ and $\bar K$. The first
term of the range term of eq. (\ref{lagd}) is
repulsive whereas the second term has an attractive contribution
for the isospin symmetric matter \cite{kmeson1} for both kaons
and antikaons.

The contributions due to the scalar-isovector, $\delta$-field
as well as the vector-isovector $\rho$-meson, introduce
isotopic asymmetry in the K and $\bar {\rm K}$-energies.
For $\rho_n > \rho_p$, in the kaon sector, $K^+$ ($K^0$)
has negative (positive) contributions from both $\delta$
and $\rho$ mesons. The $\delta$ contribution from the
scalar exchange term is positive (negative) for $K^+$ ($K^0$), whereas
that arising from the range term has the opposite sign and dominates
over the former contribution. The contribution from the $\rho$
to the kaon (antikaon) masses at $\rho_B$=0.16 fm $^{-3}$
is large as compared to that of the $\delta$ contribution,
as can be seen from the figures 2 and 3.
When we do not account for the isospin asymmetry effects arising
due to the $\rho$ and $\delta$ fields, then
the masses of kaons and antikaons stay almost constant.
The small fluctuations are reflections of the deviation of the scalar
density  occurring in the last term of equation
(\ref{selfk})((\ref{selfkb})) for the kaon (antikaon)
self energy from the baryon density, $\rho_B$.

In figure 4, the energies of the $K^+$ and $K^0$, relative to the t=0
values, are plotted for different values of the isospin asymmetry
parameter, t, at various densities. For $\rho_B$=0.16 f$m^{-3}$,
the energy of $K^+$ is seen to drop by about 15 MeV at zero momentum
when t changes from 0 to 1. On the other hand, the $K^0$ energy
is seen to increase by a similar amount for t=1, from the isospin
symmetric case of t=0.
The energies of the kaons are also plotted for densities $\rho_B$=0.31 fm$^{-3}$
and  $\rho_B$=0.61 fm$^{-3}$ in the same figure. For $K^+$, the
t-dependence of the energy is seen to be less sensitive at higher densities,
whereas the energy of $K^0$  is seen to have a larger drop from the
t=0 case, as we increase the density. The reason for this opposite
behavior for the  $K^+$ and $K^0$ on the isospin asymmetry comes
from the relative vector-meson contributions.
For $K^+$, the isospin asymmetric effect
to the energy arising from the $\rho$ meson (which dominates
over that from the $\delta$ meson) is opposite in sign to that of
the $\omega$ meson, whereas for $K^0$, it has the same sign
as that of the $\omega$ meson.

For the antikaons, the $K^-(\bar {K^0})$ energy is seen to increase
(drop) with t, as seen in figure 5. The sensitivity
of  the isospin asymmetry  dependence of the energies
is seen to be larger for $K^-$ with density, whereas it becomes
smaller for $\bar {K^0}$ at high densities. However, $\bar {K^0}$
shows an appreciable drop as we increase the momentum,
whereas the value for $K^-$ is not as sensitive to momentum
change as ${\bar K}^0$.

The qualitative behavior of the isospin asymmetry
dependencies of the energies of the kaons and antikaons
are also reflected in their optical potentials plotted
in figure 6 for the kaons, and in figure 7,
for the antikaons, at selected densities.
The different behavior of the $K^+$ and $K^0$, as well as
for the $K^-$ and $\bar {K^0}$ optical potentials
in the dense asymmetric nuclear matter should be seen
in their production as well as propagation
in isospin asymmetric heavy ion collisions.
The effects of the isospin asymmetric optical potentials
could thus be observed in nuclear collisions
at the CBM experiment at the proposed project FAIR
at GSI, where experiments with neutron rich beams
are planned to be adopted.

\section{Summary}

To summarize, we have investigated, within a chiral SU(3) model, the
density dependence of the $K, \bar K$-meson optical potentials
in asymmetric nuclear matter,
arising from the interactions with  nucleons and scalar and vector
mesons. The properties of the light hadrons -- as studied in a SU(3)
chiral model -- modify the $K (\bar K)$-meson properties in the
hadronic medium. The model with parameters fixed from the
properties of hadron masses, nuclei and KN scattering
data, takes into account all terms up to the next to leading order
arising in chiral perturbative expansion for the interactions of
$K (\bar K)$-mesons with baryons. One can observe a significant density
dependence of the isospin asymmetry on the optical potentials
of the kaons and antikaons. The results can be used in heavy-ion
simulations that include mean fields for the propagation of mesons
\cite{kmeson1}. The different potentials of kaons and antikaons can be
particularly relevant for neutron-rich beams in the CBM experiment
at the future facility FAIR at GSI, Germany,
as well as at the experiments at the planned Rare Isotope Accelerator (RIA)
laboratory, USA.

\begin{acknowledgements}

We thank Q. Li, E. Bratkovskaya, A. Dutt Majumdar, S. Sarkar, P. Roy
for many fruitful discussions.
One of the authors (AM) is grateful to the Institut
f\"ur Theoretische Physik Frankfurt for the
warm hospitality where the present work was initiated.
AM acknowledges financial support from
Alexander von Humboldt stiftung. The use of the resources of the
Frankfurt Center for Scientific Computing (CSC) is additionally
gratefully acknowledged.
\end{acknowledgements}

\end{document}